\documentclass[12pt]{article}
\usepackage{mathptmx} 
\usepackage[margin=1in]{geometry}
\usepackage{setspace}
\usepackage[sectionbib]{natbib}
\usepackage{graphicx}
\usepackage{hyperref}
\usepackage{amsmath}
\usepackage{amssymb}
\usepackage{amsfonts}
\usepackage{multirow}
\usepackage{amsthm}
\usepackage{appendix}

\theoremstyle{definition}

\usepackage[symbol,bottom]{footmisc}

\usepackage{longtable}
\usepackage{array}
\newcolumntype{P}[1]{>{\raggedright\let\newline\\\arraybackslash\hspace{0pt}}p{#1}}
\usepackage{subcaption}
\usepackage{color}
\newcommand{\widesim}[2][1.5]{
  \mathrel{\overset{#2}{\scalebox{#1}[1]{$\sim$}}}
}

\allowdisplaybreaks

\title{Comparing the Robustness of Simple \\ Network Scale-Up Method (NSUM) Estimators}
\author{Jessica P. Kunke$^1$, Ian Laga$^2$, Xiaoyue Niu$^3$, Tyler H. McCormick$^1$\\ \\
\it $^1$University of Washington, $^2$Montana State University,\\ \it $^3$Pennsylvania State University}

\begin{document}

\maketitle
\doublespace

\section{Introduction}

Surveys are a standard approach to estimating the size of subpopulations, or groups of people with a particular trait.  Many key subpopulations of interest, however, are hard to reach\footnote{We note that researchers and community partners have raised concerns about the term ``hard-to-reach"; we use this term here to encompass communities that match one or more of the above conditions, and we echo the message of \cite{routen2022HTR} that the onus of reaching such populations is on researchers and funders, not on the populations themselves.} with standard surveys for a number of possible reasons: (1) it may be hard to get a list of the members of this subpopulation or they may be hard to contact, as in the case of the homeless subpopulation; (2) it may be hard to accurately determine their membership in the subpopulation because their group status is stigmatized, as in the case of heavy drug users; and/or (3) the subpopulation may be rare in the general population \citep{Bernard1991-originalNSUM, KillworthMcCarty1998a}. The limitations of standard survey methodology in reaching these groups has motivated the development of better adapted methods.

There are generally two strategies to this estimation problem that use social networks. One is to interview an initial sample, such as a convenience sample from the hard-to-reach subpopulation of interest, then incentivize those respondents to recruit additional respondents from that subpopulation. These network sampling methods include respondent driven sampling (RDS) and snowball sampling and are often broadly called link-tracing or chain-referral methods \citep{SalganikHeckathorn2004-RDS, Handcock2014-RDS, Crawford2018-RDS}. Since these methods involve surveying members of the population of interest, they have the advantage that researchers may ask additional questions to study other aspects of the population in addition to estimating prevalence. For example, one could not only estimate the number of people who have been trafficked in a given region but also study how they entered trafficking, what enabled them to leave if they left, and what factors increased or decreased their vulnerability to trafficking. However, this approach is not always feasible, and when it is, it can be expensive, particularly if the survey aims to estimate sizes for multiple subpopulations.

The other strategy, the network scale-up method (NSUM), is to conduct a traditional survey with a representative sample from the general or frame population and ask respondents how many people they know in the hard-to-reach populations of interest, then use information about their personal network sizes (degrees) to scale up their data into an estimate for the general prevalence of those populations \citep{Bernard1991-originalNSUM, Bernard2010-nsum}. Researchers could ask respondents how many people they know (their degree) and how many people they know in the hard-to-reach group.  However, people often cannot provide an accurate estimate for their degree \citep{Killworth2003,McCormick2010-ARD}, leading to inaccurate hard-to-reach population prevalence estimates.  Therefore, respondents' degrees are often estimated by asking other questions that are used to estimate the degree.  In the probe group method, respondents are asked how many people they know in various groups of known size that collectively are thought to be representative of the general population \citep{KillworthMcCarty1998a,McCarty2001comparing,McCormick2010-ARD}. The responses to questions of the form ``How many people do you know with X trait?" are known as aggregated relational data or ARD \citep{McCormickZheng2015-ARD}. This approach does not require knowing whether the respondents themselves are in the hard-to-reach populations.

In this paper we focus on the NSUM strategy. A recent literature review by \cite{Ocagli2021} suggests the majority of studies use one of the simplest estimators. However, we find that when a key model assumption is violated by the presence of barrier effects, which may often occur in practice, the mean squared error (MSE) may actually tend to be much smaller for another simple estimator that is equally easy to implement but much less commonly used. We demonstrate this through theoretical derivations as well as simulated surveys on a real network data set.

The paper is structured as follows: Section \ref{section:nsum-background} introduces the key estimators of interest in this study. Section \ref{section:framework} presents our framework for studying the behavior of these estimators in the presence of barrier effects. Section \ref{section:analytical-results} details the analytical results comparing estimator bias, variance, and RMSE in this setting. Section \ref{section:facebook-example} compares the performance of the different estimators on surveys simulated from real network data, the Facebook 100 data set. We have shared code for the analytical and simulation results on GitHub; for review, we have omitted the URL here to deidentify the manuscript for review, but we have submitted the code as supplementary files. Section \ref{section:discussion} concludes with a discussion.

\section{NSUM estimators}
\label{section:nsum-background}

In this section, we introduce the estimators of interest in this study as compositions of degree and prevalence estimators. We discuss the assumptions underlying the models that led to these estimators, which estimators are currently used in practice, and why we reevaluate which estimator if any should be the standard.

The target of estimation is the prevalence $r$ of the hard-to-reach group $H$ in the general population, which equals the ratio of the group size $N_H$ to the size $N$ of the general or frame population.  For each respondent in a sample of size $n$, let $Y_{iH}$ represent respondent $i$'s response to the ARD question, ``How many people do you know in $H$?" Let $D_i$ represent the degree of respondent $i$. In principle, we could determine $D_i$ by directly asking respondents how many people they know in the general population, given some definition of what it means to know another person.  Since previous studies have found these responses to be inaccurate \citep{Killworth2003,McCormick2010-ARD}, researchers often estimate the degrees instead by asking other questions. Therefore, we can decompose the estimation problem into two steps: first estimating respondents' degrees, then using these degree estimates $\hat{D}_i$ with the responses $Y_{iH}$ to estimate the prevalence $r$.

The NSUM approach is based on the idea that given the response $Y_{iH}$ and degree $D_i$ for one person, a rough estimate of the hard-to-reach group prevalence is $Y_{iH}/D_i$ \citep{Bernard1991-originalNSUM}. To obtain a better estimate, we can pool the responses and degrees from a larger sample of people and model the responses $Y_{iH}$ as binomially distributed; henceforth we refer to this early NSUM model as the binomial model \citep{KillworthMcCarty1998a}. This information can be pooled in two basic ways, either the ratio of average response to average degree or the average ratio of response to degree (hereafter, we refer to the ratio of averages of a quantity as R and the average of ratios as A): \begin{equation}\hat{r}_{\text{R}} = \frac{\sum_i Y_{iH}/n}{\sum_i D_i/n} = \frac{\sum_i Y_{iH}}{\sum_i D_i}, \quad \hat{r}_{\text{A}} = \frac1n \sum_{i=1}^n \frac{Y_{iH}}{D_i}.\label{prev-estimators}\end{equation}

One approach for estimating degrees is to ask respondents how many people they know in $k$ subpopulations of known size, often called probe groups or alters. Here, known size means that the number or prevalence of each probe group in the general population can be obtained from census information or other data sources. For example, respondents might be asked how many people named Jamal they know and how many firefighters they know; this would provide responses $Y_{ij}$ for each person $i$ about probe group $j \in \{1,2\}$, where the two probe groups are the subsets of the general population (1) named Jamal and (2) serving as firefighters, respectively. Note that probe groups can and often do overlap; people named Jamal who serve as firefighters would be counted in both of these group sizes. Analogously to the prevalence estimates based on respondents' degrees, the degrees can be estimated based on these $k$ probe groups by either the R or the A degree estimator: \[\hat{D}_{i,\text{R}} = N\cdot\frac{\sum_j Y_{ij}}{\sum_j N_j}, \quad \hat{D}_{i,\text{A}} = N\cdot\frac1k \sum_{j=1}^k \frac{Y_{ij}}{N_j},\] where $N_j$ denotes the number of people in probe group $j$.

\renewcommand{\arraystretch}{2}

\begin{table}

\caption{\label{table:degree-and-prev-estimators}Each of the four hard-to-reach group prevalence estimators as functions of the survey data can be viewed as a two-step estimator, a composition of a degree estimator and a prevalence estimator that is a function of the estimated degrees. Each component estimator may be the ratio of averages (R) or the average of ratios (A). We add an asterisk to the RR$^*$ estimator here and in the text to indicate that it is the current default of the four estimators. We list MLE in quotes to indicate that the RR$^*$ estimator is not itself an MLE; the R prevalence estimator is the MLE for $r$ conditional on degrees, and the RR$^*$ uses this prevalence estimator with estimated degrees.}
\begin{center}
\begin{tabular}{ |c|c|cc|cc|c|} 
 \hline
 Name & Symbol & \multicolumn{2}{c|}{Degree step} & \multicolumn{2}{c|}{Prevalence step} & Other names\\
 \hline
 RR$^*$ & $\hat{r}_\text{RR}$ & R & $\hat{D}_i = N \cdot \frac{\sum_j Y_{ij}}{\sum_j N_j}$ & R & $\hat{r} = \frac{\sum_i Y_{iH}}{\sum_i \hat{D}_i}$ & ``MLE"  \\
 RA & $\hat{r}_\text{RA}$ & R & $\hat{D}_i = N \cdot \frac{\sum_j Y_{ij}}{\sum_j N_j}$ & A & $\:\:\hat{r} = \frac1n \sum_i \frac{Y_{iH}}{\hat{D}_i}$ & PIMLE \\
 AA & $\hat{r}_\text{AA}$ & A & $\:\:\hat{D}_i = N \cdot \frac1k \sum_j \frac{Y_{ij}}{N_j}$ & A & $\:\:\hat{r} = \frac1n \sum_i \frac{Y_{iH}}{\hat{D}_i}$ & MoS \\
 AR & $\hat{r}_\text{AR}$ & A & $\:\:\hat{D}_i = N \cdot \frac1k \sum_j \frac{Y_{ij}}{N_j}$ & R & $\hat{r} = \frac{\sum_i Y_{iH}}{\sum_i \hat{D}_i}$ & (Unmentioned) \\
 \hline
\end{tabular}
\end{center}
\end{table}

These separate degree and prevalence estimators can be combined in four ways to obtain a two-step prevalence estimator that incorporates the degree estimation step (see Table \ref{table:degree-and-prev-estimators}). To be explicit and concise, we will refer to these estimators by the choice of degree estimator followed by the choice of prevalence estimator; for example, we will use the name RA to refer to the estimator which plugs the ratio of averages for the degree estimates into the average of ratios for the prevalence estimates.

The NSUM is cost effective but also depends on fairly strong assumptions that are likely violated in many settings. The assumption that any two people are equally likely to know each other as any other two people is known as the \textbf{random mixing assumption}. The binomial model also assumes perfect visibility (each respondent knows whether each person in their network is in $H$), perfect recall (each respondent can enumerate everyone they know, or at least report the correct total), truthful answers, and the absence of other survey and response error.  In this paper, we focus specifically on the random mixing assumption.

Violations of the random mixing assumption are called \textbf{barrier effects} and are often expected in practice. Typically, we expect that not everyone in the population is equally likely to know people in the hard-to-reach population of interest. Homophily often drives connections, and people who are more similar to people in the hard-to-reach population may be more likely to know them \citep{McPherson2001_feather}. Additionally, if people in hard-to-reach subpopulations tend to have smaller or larger degrees than people in the general population, this necessarily violates the random mixing assumption, and there is evidence that this may be true for some subpopulations of interest \citep{Shelley1995}.

In the years since the early NSUM papers, a body of research has extended this model to handle barrier effects and relax the random mixing assumption; \cite{McCormick_Oxford_chapter} and \cite{Laga2021thirty} provide detailed reviews of these methods. However, these approaches are more complex and require additional data. For example, the generalized NSUM developed by \cite{FeehanSalganik2016} can be used in the presence of barrier effects and imperfect visibility, but this approach requires sampling from the hard-to-reach population in addition to the original probability sample from the general population. For this reason, \cite{FeehanSalganik2016} also describe how correction factors can be applied to the RR$^*$ estimator if sufficient data and expert knowledge exist to estimate those factors.

A systematic literature review by \cite{Ocagli2021} of all PubMed papers from the original NSUM paper through 2020 that use NSUM ultimately includes 35 studies. Examining these papers ourselves, we found that five of them focus on methods development and another two only estimate network size, not population size. Of the remaining 28 studies, all appear to use the RR$^*$ estimator, sometimes with adjustment factors; one compares against generalized NSUM, another compares against the AA estimator, and some are not explicit about the specific estimator or procedure.  None of these studies use or consider the RA or AR estimators. The most commonly used of these four estimators therefore seems to be the RR$^*$ estimator, first proposed to our knowledge by \cite{KillworthMcCarty1998b}. Throughout this paper, we write RR$^*$ as a reminder that this is the current default basic NSUM estimator.

The goal of the present study is to systematically evaluate the comparative performance of these estimators. An earlier study by \cite{KillworthMcCarty1998a} compares the RR$^*$ and RA estimators, and a more recent paper by \cite{Habecker2015-MoS} promotes the use of the AA estimator, which they call the mean-of-sums (MoS) estimator. \cite{Habecker2015-MoS} suggest potential motivations for using the AA estimator and demonstrate on an example data set that the RR$^*$ and AA estimators (as well as other modifications they propose) result in very different estimates, but they do not provide a clear theoretical or empirical evaluation of which estimator seems to be less biased. \cite{Laga2021thirty} refer to the RA estimator as the plug-in MLE, or PIMLE. We are not aware of any literature considering or proposing the use of the AR estimator, but we evaluate it in the present study along with the other three estimators.

When the binomial model is true, both the A and R prevalence estimators with fixed or known degrees are unbiased, but the latter has a smaller variance; we provide a proof in the online supplement using reasoning about harmonic and arithmetic means. \cite{KillworthMcCarty1998a} compute and empirically evaluate the RR$^*$ and RA on a data set and state that the latter has an unacceptably high variance, but their theoretical analysis assumes the degrees are known; thus their theoretical analysis concerns only the (one-step) A and R prevalence estimators, not the (two-step) RR$^*$ and RA estimators. The degree estimation step in the RR$^*$ and RA estimators not only involves additional uses of the conditional proportion assumption to estimate the degree, but also accounts for the distribution of degrees rather than conditioning on them. Perhaps this is the reason that surveys tend to use the RR$^*$ if they use one of the simple estimators, and that researchers tend to start from the RR$^*$ when they develop methods to extend the NSUM approach and relax modeling assumptions.

\section{A framework for studying estimator behavior under barrier effects}
\label{section:framework}

Our question is, how do these estimators compare under nonrandom mixing, which may be common in practice? To investigate this, we assume a model for link formation in the general population, then consider the distribution of the estimators over simple random samples from that population without replacement. The binomial model can be viewed as an approximation to the Erd{\H o}s-R\'{e}nyi network model in which the presence or absence of a link between each pair of nodes is independently drawn from the same Bernoulli($p$) distribution (see online supplement and \cite{Cheng2020-finite-sets} for details). In this study, we generalize this model to incorporate barrier effects by using a stochastic block model (SBM) that partitions the general population into two mutually exclusive groups, the hard-to-reach group of interest ($H$) and everyone else ($L$).

Under the two-group SBM, the probability of a link between any two nodes takes one of three values depending on the membership of the two nodes involved, and links are assumed to form independently from one another conditional on group membership. We denote the within-group probabilities by $p_{HH}$ and $p_{LL}$ and the between-group probability by $p_{HL}=p_{LH}$. This model will have barrier effects as long as $p_{HH} > p_{HL}$, since in that case members of $H$ will be more connected than members of $L$ to people in $H$. The dissortative condition $p_{HH} < p_{HL}$ would also create barrier effects but is typically less realistic in practice. The Erd{\H o}s-R\'{e}nyi model is a special case of this SBM with $p_{HH} = p_{HL} = p_{LH} = p_{LL}$. Note that whether the link probabilities are constant as a function of $N$ (as under the constant density assumption, so that the average degree grows linearly with total network size $N$) or scaled by $1/N$ (so that average degree is constant with respect to $N$), the expressions derived in this section will not change because the additional factors of $N$ cancel.

Many network models can be approximated by a stochastic block model with some number of blocks, perhaps many \citep{Olhede2014_universalSBM}, so the two-group SBM is a motivating choice for studying the impact of barrier effects. The two-group SBM may provide insights that have more general relevance, such as behavior based on network assortativity, even if a given problem is not believed to follow a two-group SBM.

For the sake of interpretability, we start with a simple case using estimated degrees: we suppose that respondents' degrees are estimated using one probe group $K$. Furthermore we suppose that $K \subset L$ to avoid having to specify the prevalence of $H$ within $K$, which would be necessary to compute expectations and variances. This renders the probe group unrepresentative of the general population, since it contains no one in the hard-to-reach group $H$, but the representativeness of probe groups can be a concern in practice and is therefore relevant to consider here \citep{McCormick2010-ARD}. Additionally, the real data example in Section \ref{section:facebook-example} assumes more than one probe group and does not require the probe group to be disjoint from $H$.

For the case of a single probe group, the A and R degree estimators are identical; therefore the AA and RA estimators are equivalent when degrees are estimated using a single probe group, and our analytical results will only compare the choice of prevalence estimator. Similarly, the AR and RR$^*$ estimators are equivalent when degrees are estimated using a single probe group. However, this analysis still accounts for the additional use of the random mixing assumption in estimating the degrees, and it also accounts for the distribution of degrees instead of conditioning on them. The degree estimators are no longer identical when more than one probe group is used; therefore we save comparison with the AA and AR estimators for Section \ref{section:facebook-example}.

\cite{KillworthMcCarty1998a} assume a simple random sample without replacement, while \cite{Feehan2015Thesis} and \cite{Habecker2015-MoS} propose a way to extend this to general probability survey designs. For simplicity and to stay consistent with the original estimators, in this study we assume simple random sampling without replacement.

\section{Analytical results on estimator bias and variance}
\label{section:analytical-results}

We begin by deriving approximations to the bias and variance of each estimator under a two-group stochastic block model assuming degrees are estimated using a single probe group $K$ contained in $L$.  Then to facilitate interpretation, we restrict further to the case in which the within-group link probabilities are equal to the same scaling factor $a$ times the between-group link probability $p_{HL}$. We present closed-form approximations for the bias and variance, and we numerically compute the bias, variance, and RMSE for the two estimators over a range of parameter values to characterize the regions in which one estimator outperforms the other. As mentioned previously, the AA and RA estimators are equivalent when degrees are estimated using a single probe group, as are the AR and RR$^*$ estimators; hence this section discusses only the RR$^*$ and the RA.  However, the four estimators are distinct once there is more than one probe group, so we compare all four estimators in Section \ref{section:facebook-example}.

Under the two-group SBM, the number of people person $i$ knows in the hard-to-reach group $H$ and probe group $K$, respectively, is given by \[Y_{iH} = \sum_{j=1}^{N_H^*} A_{ij} \sim \text{Binom}(N_H^*, p_{H\:g_i}), \quad Y_{iK} = \sum_{j=1}^{N_K^*} A_{ij} \sim \text{Binom}(N_K^*, p_{g_i\:L}),\] where $A$ is the adjacency matrix of the general population network; $g_i \in \{H, L\}$ denotes the group membership of person $i$; $Y_{iH}$ and $Y_{iK}$ are independent for any $i,j$; $N_H^* = N_H-1$ if $i\in H$ and $N_H^* = N_H$ otherwise; and $N_K^*$ is defined analogously to $N_H^*$. Henceforth we assume $N_H$ and $N_K$ are sufficiently large such that $N_H^* \approx N_H$ and $N_K^* \approx N_K$.

Using first-order Taylor approximations for the expectation and variance of ratios, we can estimate the expectation and variance of the estimators over the SBM superpopulation for a given sample. Since the link probability depends on the respondent's group membership, the expressions for the expectation and variance depend on $n_H$, the number of sample respondents that belong to the hard-to-reach group $H$. We then take the limit $n_H/n \to r$ as in simple random sampling without replacement (see the online supplement for further details of the derivations). In each case, the expectation is a function only of $r$ and the three link probabilities, while the variance is also a function of $n$, $N$, and $N_K$:

\vspace{-1em}
\begin{align*}
    E\left(\hat{r}_\text{RR} \right) &\to r \frac{ r p_{HH} + (1-r) p_{HL}}{r p_{HL} + (1-r) p_{LL}} \\
    E\left(\hat{r}_\text{RA} \right) &\to r \left[ r \frac{p_{HH}}{p_{HL}} + (1-r) \frac{p_{HL}}{p_{LL}} \right] \\
    \text{Var}\left(\hat{r}_\text{RR} \right) &\to \frac{r}{nN} \frac{\left(r p_{HL} + (1-r) p_{LL} \right)^2 \left[ r p_{HH} (1-p_{HH}) + (1-r) p_{HL}(1-p_{HL})\right]}{\left(r p_{HL} + (1-r) p_{LL} \right)^4} \:+\: \\
    &\quad \frac{r^2}{nN_K}\frac{ \left(r p_{HH} + (1-r) p_{HL} \right)^2 \left[ r p_{HL} (1-p_{HL}) + (1-r) p_{LL}(1-p_{LL})\right]}{\left(r p_{HL} + (1-r) p_{LL} \right)^4} \\
    \text{Var}\left(\hat{r}_\text{RA} \right) &\to \frac{r}{n N p_{HL}^2} \big[ r p_{HH}(1-p_{HH}) + (1-r) p_{LL}(1-p_{HL}) \big] \:+ \\
    &\quad \frac{r^2}{n N_K p_{HL}^3} \big[ r p_{HH}^2(1-p_{HL}) + (1-r) p_{HL}^2(1-p_{LL}) \big]
\end{align*}

Note that when the three link probabilities are equal, corresponding to the binomial model, the RR$^*$ estimator does not have smaller variance as suggested by \cite{KillworthMcCarty1998a}: both estimators have the same first-order variance. We believe this is the first time this result has been shown for these estimators. In this case, the first-order approximations of both estimators' expectations equal the true prevalence; in the language of \cite{FeehanSalganik2016} and other literature, both the RR$^*$ and RA estimators are essentially unbiased.

Thus we have expressions for the bias and variance of each estimator under a general two-group stochastic block model when degrees are estimated using a single probe group $K\subset L$.  These expressions describe how the expectation and variance depend on various parameters such as $r$ and the link probabilities. However, we would like to be able to characterize the regions of parameter space in which each estimator performs better than the other: under which combinations of parameters is one estimator better than the other?  With four parameters for the expectation and seven for the variance, it is hard to identify patterns that summarize when to use which estimator. Therefore we now analyze a slightly simpler case in which we further specify the relationships among the three link probabilities: Fix $p_{HH} = p_{LL} = ap_{HL}$ for some $0 < a < \infty$. Notice that $a=1$ corresponds to the Erd{\H o}s-R\'{e}nyi case, $a>1$ corresponds to assortativity, and $a<1$ corresponds to dissortativity. With these additional constraints, we can characterize the three link probabilities with just the two parameters $a$ and $p_{HL}$, the latter of which we will now denote simply by $p$. This reduces the number of degrees of freedom by two, so that the expectation is now a function of just two parameters and the variance is a function of five.

The biases are now a function only of $a$ and $r$:
\begin{align*}
    \text{Bias}\left(\hat{r}_\text{RR} \right)(a, r) &\to r\left[\frac{(a-1)(2r-1)}{(1-r)a + r}\right] \\
    &= \begin{cases}
     >0 & \{a>1\} \cap \{r > 0.5\} \:\text{ or }\: \{a<1\} \cap \{r < 0.5\}, \\
     =0 & \{a=1\} \cup \{r = 0.5\}, \\
     <0 & \text{else},
    \end{cases} \\
    \text{Bias}\left(\hat{r}_\text{RA} \right)(a, r) &\to r \left[ \frac{(a-1)\big[(a+1)r-1\big]}{a} \right] \\
    &= \begin{cases}
     >0 & \{a>1\} \cap \{a > \frac{1-r}{r}\} \:\text{ or }\: \{a<1\} \cap \{a < \frac{1-r}{r}\}, \\
     =0 & \{a=1\} \cup \{a = \frac{1-r}{r}\}, \\
     <0 & \text{else}.
    \end{cases} \\
\end{align*}
\vspace{-5em}

The RR$^*$ is unbiased if and only if the Erd{\H o}s-R\'{e}nyi case holds ($a=1$) or the hard-to-reach population prevalence is exactly 50\%. The RA is unbiased if and only if the Erd{\H o}s-R\'{e}nyi case holds ($a=1$) or $a = (1-r)/r$. It is unlikely that the parameters take these exact values such that the estimators are exactly unbiased, but these conditions serve as boundary cases to define regions in which one estimator or the other has smaller bias.

\begin{figure}
    \centering
    \includegraphics[width=\textwidth]{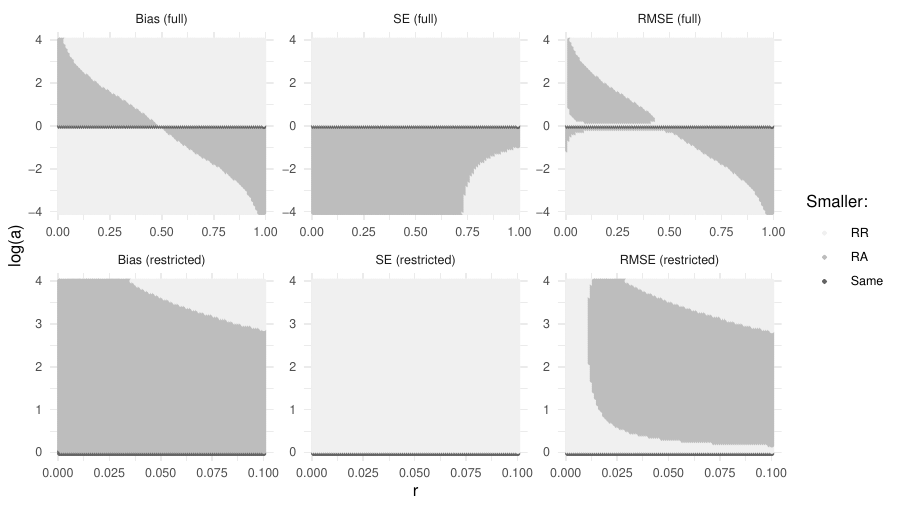}
    \caption{Comparing estimator bias (left panel), variance (center panel), and RMSE (right panel) as a function of $a$ and $r$. The darkest regions indicate the combinations of $a$ and $r$ for which the RA has the lower value of the quantity examined in that subplot. Note the log scale (base-$e$) on $a$, such that the x-axis corresponds to the Erd{\H o}s-R\'{e}nyi case $a=1$ in which the two estimators have the same bias and variance. Here, $p=0.01$, $n\cdot N = 500,000$, and $r_K := N_K/N = 0.1$. The top row shows results over a wider range of parameter values while the bottom row shows results for a smaller range of parameter values thought to reflect most practical settings, namely $a>1$ (assortative) and prevalence smaller than 10\%. Under assortativity and for prevalences less than 10 or 20\%, the RA often has smaller bias and RMSE than the RR$^*$, and the choice of estimator should depend on the specific populations and study.}
    \label{compare-bias-rmse-analytical}
\end{figure}

Now we approximate the variances. Denoting the prevalence of the probe group $K$ by $r_K := N_K/N$, the dependence simplifies to effectively five parameters: $r$, $a$, $p$, $r_K$, and $nN$, since the dependence on sample size $n$ and population size $N$ is only through their product.

\vspace{-1em}
\begin{align*}
    \text{Var}\left(\hat{r}_\text{RR} \right) &\to \frac{r}{nNp} \bigg[\frac{\left(r + (1-r)a \right)^2 \left( ra + (1-r) - \left[ra^2 + (1-r)\right] p\right)}{\left[r + (1-r) a \right]^4} \:+\: \\
    &\qquad\qquad \frac{r}{r_K}\frac{\left(ra + (1-r) \right)^2 \left( r + (1-r)a - \left[r + (1-r)a^2\right] p\right)}{\left[r + (1-r) a \right]^4}\bigg] \\
    \text{Var}\left(\hat{r}_\text{RA} \right) &\to \frac{r}{n N p} \left(pa(1-a)r + (1-p)a + \frac1{r_K} \bigg[\big([1-p]a^2 + pa - 1\big) r^2 + (1-pa)r\bigg]\right)
\end{align*}

To characterize the conditions under which each estimator outperforms the other, we evaluate the approximate bias and variance expressions above over a range of parameter values and visualize the regions in which each estimator has lower bias, lower variance, and lower RMSE. In the top row of Figure \ref{compare-bias-rmse-analytical} we show the results for $\log(a)$ ranging from -4 to 4 in increments of 0.1, and $r$ ranging from 0.01 to 0.99 in increments of 0.02. The bottom row shows the results restricted to assortative cases with small $r$, with $\log(a)$ ranging from 0 to 4 in increments of 0.05, and $r$ ranging from 0.001 to 0.1 in increments of 0.001. In both these cases we fix $p=0.01$, $nN = 500,000$, and $r_K = 0.1$ since the variance changes with these parameters.

The leftmost panels of Figure \ref{compare-bias-rmse-analytical} compare the bias of the two estimators as a function of $a$ and $r$; the darkest shaded region is the region in which the RA has smaller bias in magnitude. We have already shown the two estimators have the same bias and variance and therefore the same RMSE for $a=0$, which is the x-axis in these figures. In practice, we typically expect assortativity ($a>1$) and a fairly small value of $r$, probably less than 10 or 20\%; throughout most of this region of the parameter space (for any values of the other parameters, since the estimator biases depend only on $a$ and $r$), the RA has smaller bias.

The middle panels of Figure \ref{compare-bias-rmse-analytical} illustrate that the variance is smaller for the RR$^*$ under assortativity and generally smaller for the RA under dissortativity. The rightmost panels comparing the RMSE of the two estimators resemble the leftmost panels comparing the bias. Therefore, in the settings most likely to be practically relevant, and for sample sizes and population sizes likely to be realistic, the RA tends to have smaller RMSE. For assortative settings with small prevalence $r$, the RA has lower RMSE.  When the assortativity is weaker, the RA has lower RMSE for a wider range of $r$, and under stronger assortativity the upper bound on $r$ for this region shrinks.

Therefore, although the RR$^*$ is the most commonly used estimator in the literature, the RA may often have lower bias and RMSE in the presence of barrier effects.

The relative importance of the bias and variance in the RMSE are determined by $nN$, $r_K$, and $p$. The region of parameter space in which the RA has lower RMSE expands with $nN$ and with $p$ and shrinks with $r_K$; see the online supplement for figures and additional details. The size of this region depends more strongly on $nN$ and $p$ than on $r_K$.

\section{Facebook 100 data example}
\label{section:facebook-example}

We also conduct simulations using an example data set, and the results are consistent with the analytical results in Section \ref{section:analytical-results} even though (1) these networks are not simulated from a two-group stochastic block model and (2) we use multiple probe groups to estimate the degrees.

For these simulations we use the Facebook 100 data set, which consists of the intra-school links in the September 2005 Facebook networks of 100 colleges and universities \citep{facebook, facebooklong}. The networks range in size from 769 to 41,554 nodes, and 91\% of the schools have fewer than 25,000 nodes. Similar to \cite{Feehan2022_Facebook_weakties}, we create candidate probe groups from the following five variables: status (such as faculty or student), gender, year, dorm, and major. We treat them as categorical variables, with an indicator for each level of each variable, and each indicator variable whose prevalence in that school network is 0.1-10\% of the population is a candidate probe group for that school network. In other words, if people living in Dorm 18 at a given school represent 0.1-10\% of that school's Facebook network, we include this group in the list of candidate probe groups for that school.

We simulate surveys on this complete network data set, and since we know the true prevalence of each group in the networks, we can compare the bias, standard error (SE), and root mean square error (RMSE) of the four estimators. Some covariate data is missing in this data set, but the networks as they are recorded are taken to be the true networks for the purpose of evaluating the estimators in this study. For example, the number of people in a given major is taken to be the number of people whose major is recorded as that value.

For each school network, we select the 20 largest candidate probe groups with low assortativity (we require the assortativity coefficient magnitude to be less than 0.1) to constitute the probe groups used in estimating respondents' degrees. From the remaining candidate groups, we choose the 10 most assortative groups and the 10 least assortative groups (those with the smallest or most negative assortativity coefficients) to be the hard-to-reach groups whose prevalences we will estimate; we will refer to a choice of school network and hard-to-reach group as a case. In the results and discussion presented here, we focus on the assortative cases since we believe those to be more relevant in practice, but we comment also on the low-assortativity cases. One school has only 29 candidate groups, so we only evaluate 9 cases for this school, resulting in a total of 999 assortative cases and 990 low-assortativity cases across the 100 schools. The prevalences of the resulting set of assortative hard-to-reach groups range from 0.010 to 9.9\%, 2.2\% on average.

For each case, we draw 500 ``survey" samples of 500 people each using simple random sampling without replacement to represent our survey respondents, and for each survey sample we compute the RR$^*$, RA, AR, and AA estimates for that case. We compute the responses for each survey respondent directly from the network data; for example, we compute each person's response to ``How many people do you know in group X?" as the number of people they are connected to in that group. We approximate the mean and SE of each estimator for each case as the sample mean and sample standard deviation across the 500 surveys for that case, then use this to estimate the bias and RMSE for each estimator.

We categorize these cases based on their degree ratio, computed by averaging the degrees of everyone in the hard-to-reach group and dividing by the average across the degrees of everyone in the general population; in this context, the hard-to-reach group is taken to be a subset of the general population and a person's degree is the number of people they know in whatever group is defined to be the general population \citep{FeehanSalganik2016}.  We consider three broad categories in which the degree ratio is ``low" ($<0.8$, 290 cases), ``high" ($>1.2$, 360 cases), or near 1 (between 0.8 and 1.2, 349 cases). The assortativity coefficient of each case ranges from 0.019 to 0.87, with first, second, and third quartiles of 0.23, 0.29, and 0.34.

It is worth noting that in these simulations, some degrees are estimated to be zero; overall across the 100 networks, 2.3\% of the degree estimates are zero, ranging from 0.4-5.3\% for each network. As a result, the RA and AA sample estimates for a given survey are undefined when both the numerator (response) and denominator (estimated degree) are zero for a given person, and infinite when only the estimated degree is zero. For each case, we compared four different methods of handling the zero-valued degree estimates when computing the mean and standard deviation of the survey estimates, and we found that the choice of method made little difference in the performance of the estimators.  Further details are provided in the online supplement. For the rest of the analysis presented here, we set infinite values to 1 and exclude undefined values.

Even when the estimated degrees are not zero, the ratio of response to estimated degree $Y_{iH}/\hat{D}_i$ is nonsensically greater than 1 if the response is greater than the estimated degree. This is true for 0.14\% of the people across these 100 networks. In the simulations, we have handled this by replacing any such ratios with 1; effectively, this assumes that for anyone whose estimated degree is smaller than their response, their personal network consists only of people in the hard-to-reach population. It is not that we think this is true, since the person may just know disproportionately fewer people in the probe groups than one would expect based on their prevalences, but it is one way to handle these cases in the absence of further knowledge. Nonetheless, in these 100 networks, we do find that people with greater responses than estimated degrees tend to know many more people in the hard-to-reach group than average (22 on average, versus an average of 1.2 for everyone across the 100 networks) and simultaneously have smaller true degrees than average (36 versus 78), so the distribution of their true ratios is skewed toward 1. Omitting these nonsensical ratios instead of replacing them with 1 leads to similar results as those we report below.

The degree estimates are reasonable but not idealized. The first, second and third quartiles for percent error in the degree estimates are 9.1, 19.4, 33.9\% for the R degree estimates and 9.3, 19.8, 35.5\% for the A degree estimates, respectively. Correlation between estimated and true degrees across all the networks using the chosen probe groups is 0.976 for the R estimates, 0.974 for the A estimates. Scatterplots of estimated versus true degrees are available in the online supplement.

Overall, the RA (AA, RR$^*$, AR) estimator has the lowest RMSE of the four estimators in 41\% (21\%, 20\%, 18\%) of the 999 cases. The RA has lower RMSE than the AA (RR$^*$, AR) estimator in 67\% (62\%, 62\%) of the cases (see Table \ref{table:overall-pairwise-comparisons}). The RR$^*$ estimator tends to have the lowest SE (46\% of the cases) while the AA estimator tends to have the highest (55\% of the cases). The RA estimator tends to have the lowest bias (43\% of cases) while the AR and RR$^*$ tend to have the highest (34\% of cases for AR, 33\% for RR$^*$).

\renewcommand{\arraystretch}{1.2}

\begin{table}
\caption{\label{table:overall-pairwise-comparisons} Pairwise comparisons of the four estimators by RMSE. For example, the RA estimator has smaller RMSE than the RR$^*$ estimator in 61.5\% of the cases.}
\begin{center}
\begin{tabular}{ |c|c|c|c|c|} 
 \hline
 & & \multicolumn{3}{c|}{Larger RMSE} \\
 \hline
 & & AA & RR$^*$ & AR \\
 \hline
 Smaller & RA & 67.1\% & 61.5\% & 61.9\% \\
 RMSE & AA & - & 58.4\% & 58.6\% \\
  & RR$^*$ & - & - & 52.3\% \\
 \hline
\end{tabular}
\end{center}
\end{table}

\begin{figure}
    \centering
    \includegraphics{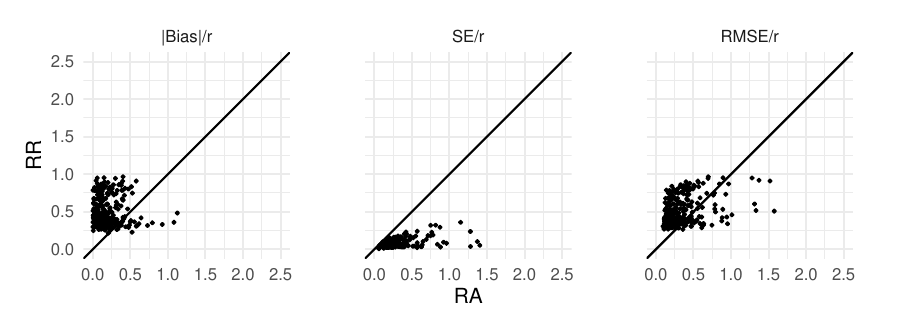}
    \caption{Comparing the absolute-value bias (left panel), SE (center panel), and RMSE (right panel), all standardized by the true prevalence, of the RR$^*$ and the RA in the 290 ``low" degree ratio cases ($<0.8$) from the Facebook 100 simulations. We standardize these metrics by the true prevalence of each case since the true prevalence varies widely across cases. Each point represents the average across 500 surveys of size 500 for one combination of school network and hard-to-reach group. The diagonal line is the one-to-one line; points above the line have lower values for the RA than the RR$^*$. Among these cases, the RA tends to have smaller bias, larger variance, and smaller RMSE than the RR$^*$ estimator.}
    \label{compare-RA-RR-fb100-low}
\end{figure}

Among cases with low degree ratios, thought to be more relevant to some hard-to-reach populations, the RA estimator has lower RMSE than the RR$^*$ (AR, AA) estimator in 80\% (80\%, 64\%) of the cases. Figure \ref{compare-RA-RR-fb100-low} illustrates that while the RA tends to have larger variance than the RR$^*$ estimator in these cases, it tends to have smaller bias and RMSE.

\begin{figure}
    \centering
    \includegraphics{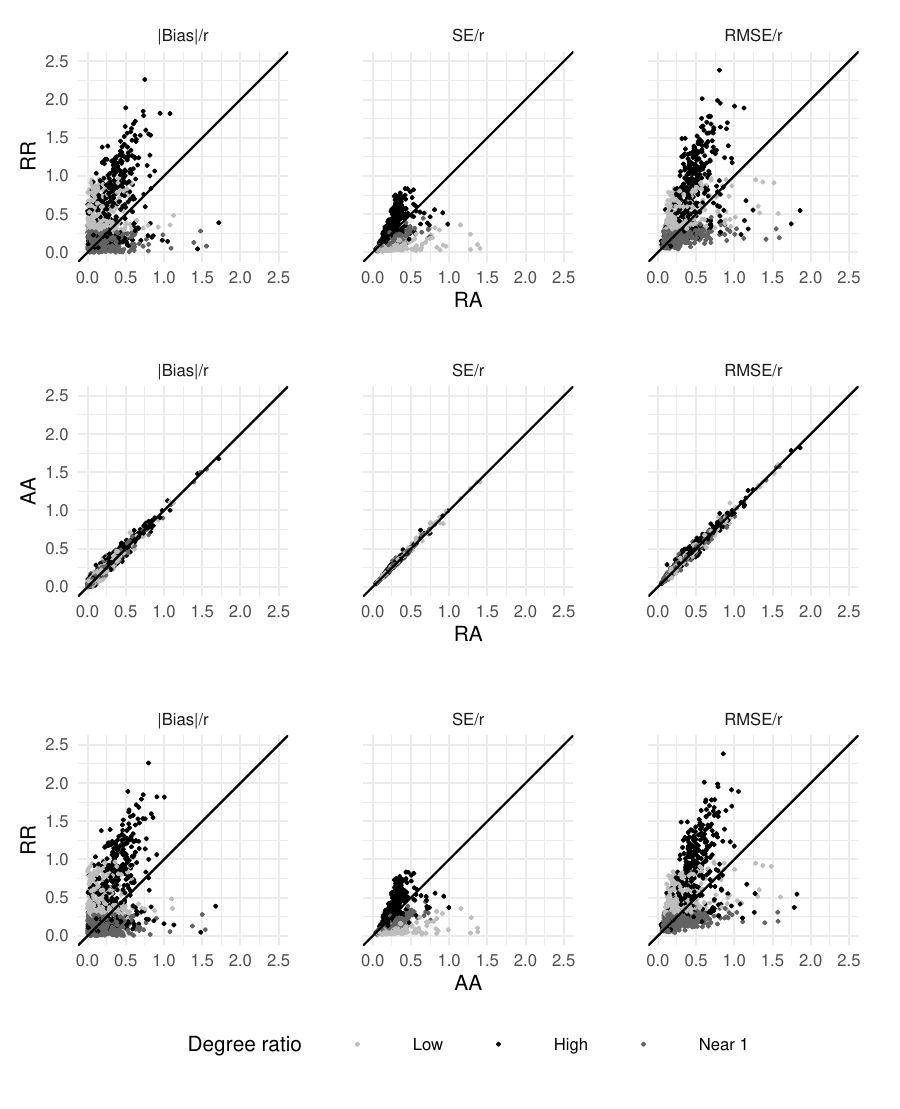}
    \caption{As in Figure \ref{compare-RA-RR-fb100-low} except comparing three of the four estimators for all 999 cases. Points representing cases with low, near-1 and high degree ratios are shown in light grey, dark grey, and black, respectively. The RA and AA estimators tend to have lower bias and RMSE than the RR$^*$ and AR estimators for cases with low and high degree ratios.}
    \label{compare-all-fb100}
\end{figure}

Figure \ref{compare-all-fb100} shows that the RA estimator tends to have lower RMSE than the RR$^*$ estimator when the degree ratio is high as well, and higher RMSE when the degree ratio is near 1. \cite{FeehanSalganik2016} demonstrate that the bias of the RR$^*$ estimator is a function of the degree ratio, increasing as the degree ratio diverges from 1. In these simulations, the RA estimator tends to have lower bias than the RR$^*$ estimator in the low- and high-degree ratio cases. Therefore, researchers that are confident the degree ratio is near 1 for the hidden and general population of interest to them may want to use the RR$^*$, but otherwise the RA may have lower RMSE.

We consider how these pairwise comparisons change with respect to various other factors. When the cases are grouped by true prevalence or by the percent of people in the network whose estimated degrees are zero, within each group the RMSE and bias still tend to be lower for the RA estimator than the other estimators. Among the low-assortativity cases, we see the opposite trend in which the RA estimator tends to have worse RMSE than the RR$^*$ estimator, consistent with the analytical results presented above.

When the cases are grouped by an estimate of $p_{HL}$, computed by dividing the number of observed $H-L$ edges by the possible number of $H-L$ edges ($N_H(N-N_H)$), within each group the RMSE and bias still tend to be lower for the RA estimator than the other estimators except for the lowest quartile of the estimated probabilities (our estimates for $p_{HL}$ range from $9.4\times 10^{-5}$ to $8.8\times 10^{-2}$, and the first quartile is as large as $3.0\times 10^{-3}$).  This agrees with our analytical result above that the region in which the RA estimator has lower RMSE shrinks as this link probability decreases.

In these Facebook 100 simulations, the RA has smaller bias, variance, and RMSE than the AA estimator in more than half the cases, even among cases with degree ratios near 1. Recall that the AA estimator uses the same size estimator as the RA but uses the average of ratios for the degree estimator as well (Table \ref{table:degree-and-prev-estimators}). The smaller variance is not surprising since the A estimator has greater variance than the R estimator as demonstrated in the online supplement. Yet this choice appears to increase not only the variance but the bias. We suspect the increased bias follows from the fact that the probe groups are intended to be collectively but not necessarily individually representative of the general population \citep{McCormick2010-ARD}. If the set of chosen probe groups satisfy this property, then taking the ratio of averages keeps the numerator and denominator representative while taking the average of ratios changes the relative weighting of the probe groups.

In the set of networks and choices of probe groups here, the A and R degree estimates are very comparable. In these simulations, the percent error is at least 0.5\% smaller for the R estimator than the A estimator in 49.8\% of the cases, and the reverse is true in 42.2\% of the cases. The distributions of their percent errors are very similar, and the two estimators are highly correlated with each other. When the two estimators do not perform so similarly, the difference in performance between the RA and AA estimators (and between the RR$^*$ and AR estimators) could be greater.

\section{Discussion}
\label{section:discussion}

We have presented theoretical evidence that over what seems likely to be a realistic range of true prevalence, sample size, and population size, the RA estimator often has lower bias and RMSE than the RR$^*$ estimator under a two-block stochastic block model with assortativity. We have also provided empirical evidence on real networks that the RA estimator may often have lower bias and RMSE than the RR$^*$ estimator under assortativity.  Together, these results suggest that using the ratio-of-averages degree estimator and the average-of-ratios size estimator often has lower bias and RMSE than the most commonly used NSUM basic scale-up estimator, which uses the ratio of averages for both the degree and size estimators.

We have also shown empirical evidence that the RA estimator often has lower bias, RMSE, and variance than the AA (mean-of-sums) estimator, suggesting that using the average-of-ratios size estimator with the ratio-of-averages degree estimator may be better than using the average-of-ratios size estimator for both the size and degree estimators. We have provided some reasoning for these findings.

Generalized NSUM and other complex methods are likely to be preferable to these simple estimators when feasible. However, in studies that use one of these simple estimators without any corrections, the RA may be the most robust choice.  When the necessary data are available, it may be easier to correct bias in the RR$^*$ estimator than the RA since the RR$^*$ bias can be expressed directly in terms of the degree ratio; however, this requires having reliable estimates of the degree ratio.  When ad hoc corrections to the simple estimators are applied to the RR$^*$, it may be helpful to also compute the RA to help bound the result; it is easily done and does not require additional data or computational power.

The theoretical and empirical findings presented here raise the question whether it may be more robust to develop methods based on the RA instead of the RR$^*$.  This may depend on how successfully we can characterize the bias of the RA and correct for it.  It may turn out that we can more readily understand and correct for the bias of the RR$^*$.

These findings do not indicate that the RA estimator is always best, but rather that it outperforms the other estimators by enough of a margin in enough potentially relevant parameter combinations that it is worth considering (1) as a default choice in the absence of further information and (2) as the basis for future method development. The evidence in favor of one estimator over another still depends on the range of parameter values expected to be practically relevant. The more we know about the typical network and sample size, assortativity, approximate prevalence, typical degree sizes, and the fraction of the population that is in one or more of the probe groups in hard-to-reach population studies, the better we can constrain the relative performance of the estimators.

Additionally, aside from rarity (low prevalence) in some of the cases, the Facebook 100 network data do not represent subpopulations we expect to be hidden, stigmatized, or hard to reach for other reasons. Comparing the robustness of these estimators on additional data sets beyond our initial example, particularly using network data sets involving hard-to-reach populations, would provide more direct evidence for evaluating these estimators.

We have interpreted the simulation results in terms of the proportion of cases in which one estimator performs better than another, or the distribution of relative estimator performance across the cases.  The set of cases is not a sample from a distribution and should be interpreted with caution.

The zero-valued degree estimates that arose in our simulations beg the question of how to handle these cases in practice. We considered four different methods, three of which reduce the effective sample size by excluding respondents with zero-valued degree estimates and non-zero responses, zero-valued degree estimates and zero-valued responses, or both (see the online supplement for details). The impact of the reduced sample size in a given study depends on how common these two types of cases are in that particular study. In the current study, the choice of method only impacts the ranking of estimators by RMSE in a handful of simulated cases, but further research is needed to study when and why that choice matters and how greatly it affects relative estimator performance.  Perhaps the fourth method which does not exclude any respondents' data is sufficient for handling zero-valued degree estimates, but it would be valuable to study how large a proportion of the respondents would have to have zero-valued degree estimates before this method introduces too much bias.  Otherwise, the reduction in effective sample size by the RA estimator may be another cost to consider when deciding which estimator to use in a given study.

In both the analytical and simulation studies in this paper, the only source of error is nonrandom mixing. In practice, of course, we expect other errors. In this study, for instance, we suppose we knew respondents' true responses; in reality, a respondent may not realize someone they know is a member of the group in question or they may not want to reveal they know people in that group, leading to underestimated responses. Other errors could tend to overestimate the response or have different effects in different contexts. Future work can examine the impact on relative estimator performance in the case that these errors impact the responses but not the degree estimates, or in the case that the errors impact both.

Combining NSUM and other methods may also prove to be a helpful strategy. Hard-to-reach populations are often studied in the context of estimating the impact of a social or epidemiological concern; for instance, estimating the number of people who have experienced labor trafficking is part of an effort to understand, intercept and prevent trafficking. Studies may benefit from using RDS and related methods to learn about possible causes and interventions, while using NSUM to do more frequent monitoring. For example, researchers and monitoring agencies could use NSUM regularly to estimate how many people are being trafficked over time, and implement less frequent RDS studies to both calibrate the NSUM estimates and learn from people who have been trafficked how they entered their situation, how they were able to leave the situation, or what prevented them from leaving.

\bibliographystyle{asr}
\bibliography{nsum}

\begin{thebibliography}{}
\newcommand{\enquote}[1]{``#1''}

\bibitem[\protect\citeauthoryear{Bernard, Hallett, Iovita, Johnsen, Lyerla,
  McCarty, Mahy, Salganik, Saliuk, Scutelniciuc, Shelley, Sirinirund, Weir, and
  Stroup}{Bernard et~al.}{2010}]{Bernard2010-nsum}
Bernard, H.~Russell, Tim Hallett, Alexandrina Iovita, Eugene~C. Johnsen, Rob
  Lyerla, Christopher McCarty, Mary Mahy, Matthew~J. Salganik, Tetiana Saliuk,
  Otilia Scutelniciuc, Gene~A. Shelley, Petchsri Sirinirund, Sharon Weir, and
  Donna~F. Stroup. 2010.
\newblock \enquote{Counting hard-to-count populations: the network scale-up
  method for public health.}
\newblock {\em Sexually Transmitted Infections\/} 86:1368--4973.

\bibitem[\protect\citeauthoryear{Bernard, Johnsen, Killworth, and
  Robinson}{Bernard et~al.}{1991}]{Bernard1991-originalNSUM}
Bernard, H.~Russell, Eugene~C. Johnsen, Peter~D. Killworth, and Scott Robinson.
  1991.
\newblock \enquote{Estimating the size of an average personal network and of an
  event subpopulation: Some empirical results.}
\newblock {\em Social Science Research\/} 20:109--121.

\bibitem[\protect\citeauthoryear{Cheng, Eck, and Crawford}{Cheng
  et~al.}{2020}]{Cheng2020-finite-sets}
Cheng, Si, Daniel~J. Eck, and Forrest~W. Crawford. 2020.
\newblock \enquote{Estimating the size of a hidden finite set: Large-sample
  behavior of estimators.}
\newblock {\em Statistics Surveys\/} 14:1--31.

\bibitem[\protect\citeauthoryear{Crawford, Wu, and Heimer}{Crawford
  et~al.}{2018}]{Crawford2018-RDS}
Crawford, Forrest~W., Jiacheng Wu, and Robert Heimer. 2018.
\newblock \enquote{Hidden Population Size Estimation From Respondent-Driven
  Sampling: A Network Approach.}
\newblock {\em Journal of the American Statistical Association\/} 113:755--766.
\newblock PMID: 30828120.

\bibitem[\protect\citeauthoryear{Feehan}{Feehan}{2015}]{Feehan2015Thesis}
Feehan, Dennis~M. 2015.
\newblock {\em Network Reporting Methods\/}.
\newblock Ph.D. thesis, Princeton University.

\bibitem[\protect\citeauthoryear{Feehan and Salganik}{Feehan and
  Salganik}{2016}]{FeehanSalganik2016}
Feehan, Dennis~M. and Matthew~J. Salganik. 2016.
\newblock \enquote{Generalizing the network scale-up method: A new estimator
  for the size of hidden populations.}
\newblock {\em Sociological Methodology\/} 46:153--186.

\bibitem[\protect\citeauthoryear{Feehan, Son, and Abdul-Quader}{Feehan
  et~al.}{2022}]{Feehan2022_Facebook_weakties}
Feehan, Dennis~M., Vo~Hai Son, and Abu Abdul-Quader. 2022.
\newblock \enquote{Survey Methods for Estimating the Size of Weak-Tie Personal
  Networks.}
\newblock {\em Sociological Methodology\/} 52:193--219.

\bibitem[\protect\citeauthoryear{Habecker, Dombrowski, and Khan}{Habecker
  et~al.}{2015}]{Habecker2015-MoS}
Habecker, Patrick, Kirk Dombrowski, and Bilal Khan. 2015.
\newblock \enquote{Improving the Network Scale-Up Estimator: Incorporating
  Means of Sums, Recursive Back Estimation, and Sampling Weights.}
\newblock {\em PLOS ONE\/} 10:1--16.

\bibitem[\protect\citeauthoryear{Handcock, Gile, and Mar}{Handcock
  et~al.}{2014}]{Handcock2014-RDS}
Handcock, Mark~S., Krista~J. Gile, and Corinne~M. Mar. 2014.
\newblock \enquote{Estimating hidden population size using Respondent-Driven
  Sampling data.}
\newblock {\em Electronic Journal of Statistics\/} 8:1491--1521.

\bibitem[\protect\citeauthoryear{Killworth, Johnsen, McCarty, Shelley, and
  Bernard}{Killworth et~al.}{1998a}]{KillworthMcCarty1998a}
Killworth, Peter~D., Eugene~C. Johnsen, Christopher McCarty, Gene~Ann Shelley,
  and H.~Russell Bernard. 1998a.
\newblock \enquote{A social network approach to estimating seroprevalence in
  the {U}nited {S}tates.}
\newblock {\em Social Networks\/} 20:23--50.

\bibitem[\protect\citeauthoryear{Killworth, McCarty, Bernard, Johnsen, Domini,
  and Shelley}{Killworth et~al.}{2003}]{Killworth2003}
Killworth, Peter~D, Christopher McCarty, H~Russell Bernard, Eugene~C Johnsen,
  John Domini, and Gene~A Shelley. 2003.
\newblock \enquote{Two interpretations of reports of knowledge of subpopulation
  sizes.}
\newblock {\em Social networks\/} 25:141--160.

\bibitem[\protect\citeauthoryear{Killworth, McCarty, Bernard, Shelley, and
  Johnsen}{Killworth et~al.}{1998b}]{KillworthMcCarty1998b}
Killworth, Peter~D., Christopher McCarty, H.~Russell Bernard, Gene~Ann Shelley,
  and Eugene~C. Johnsen. 1998b.
\newblock \enquote{Estimation of seroprevalence, rape, and homelessness in the
  {U}nited {S}tates using a social network approach.}
\newblock {\em Evaluation Review\/} 22:289--308.

\bibitem[\protect\citeauthoryear{Laga, Bao, and Niu}{Laga
  et~al.}{2021}]{Laga2021thirty}
Laga, Ian, Le~Bao, and Xiaoyue Niu. 2021.
\newblock \enquote{Thirty Years of the Network Scale-up Method.}
\newblock {\em Journal of the American Statistical Association\/}
  116:1548--1559.

\bibitem[\protect\citeauthoryear{McCarty, Killworth, Bernard, Johnsen, and
  Shelley}{McCarty et~al.}{2001}]{McCarty2001comparing}
McCarty, Christopher, Peter~D Killworth, H~Russell Bernard, Eugene~C Johnsen,
  and Gene~A Shelley. 2001.
\newblock \enquote{Comparing two methods for estimating network size.}
\newblock {\em Human organization\/} 60:28--39.

\bibitem[\protect\citeauthoryear{McCormick}{McCormick}{2021}]{McCormick_Oxford_chapter}
McCormick, Tyler~H. 2021.
\newblock \enquote{{The Network Scale-Up Method}.}
\newblock In {\em {The Oxford Handbook of Social Networks}\/}. Oxford
  University Press.

\bibitem[\protect\citeauthoryear{McCormick, Salganik, and Zheng}{McCormick
  et~al.}{2010}]{McCormick2010-ARD}
McCormick, Tyler~H., Matthew~J. Salganik, and Tian Zheng. 2010.
\newblock \enquote{How Many People Do You Know?: Efficiently Estimating
  Personal Network Size.}
\newblock {\em Journal of the American Statistical Association\/} 105:59--70.
\newblock PMID: 23729943.

\bibitem[\protect\citeauthoryear{McCormick and Zheng}{McCormick and
  Zheng}{2015}]{McCormickZheng2015-ARD}
McCormick, Tyler~H. and Tian Zheng. 2015.
\newblock \enquote{Latent Surface Models for Networks Using Aggregated
  Relational Data.}
\newblock {\em Journal of the American Statistical Association\/}
  110:1684--1695.

\bibitem[\protect\citeauthoryear{McPherson, Smith-Lovin, and Cook}{McPherson
  et~al.}{2001}]{McPherson2001_feather}
McPherson, Miller, Lynn Smith-Lovin, and James~M Cook. 2001.
\newblock \enquote{Birds of a Feather: Homophily in Social Networks.}
\newblock {\em Annual Review of Sociology\/} 27:415--444.

\bibitem[\protect\citeauthoryear{Ocagli, Azzolina, Lorenzoni, Gallipoli,
  Martinato, Acar, Berchialla, Gregori, Group, et~al.}{Ocagli
  et~al.}{2021}]{Ocagli2021}
Ocagli, Honoria, Danila Azzolina, Giulia Lorenzoni, Silvia Gallipoli, Matteo
  Martinato, Aslihan~S Acar, Paola Berchialla, Dario Gregori, INCIDENT~Study
  Group, et~al. 2021.
\newblock \enquote{Using Social Networks to Estimate the Number of {COVID}-19
  Cases: The {INCIDENT (Hidden COVID-19 Cases Network Estimation)} Study
  Protocol.}
\newblock {\em International Journal of Environmental Research and Public
  Health\/} 18:5713.

\bibitem[\protect\citeauthoryear{Olhede and Wolfe}{Olhede and
  Wolfe}{2014}]{Olhede2014_universalSBM}
Olhede, Sofia~C. and Patrick~J. Wolfe. 2014.
\newblock \enquote{Network histograms and universality of blockmodel
  approximation.}
\newblock {\em Proceedings of the National Academy of Sciences\/}
  111:14722--14727.

\bibitem[\protect\citeauthoryear{Routen, Bambra, Willis, and Khunti}{Routen
  et~al.}{2022}]{routen2022HTR}
Routen, A, C~Bambra, A~Willis, and K~Khunti. 2022.
\newblock \enquote{Hard to reach? Language matters when describing populations
  underserved by health and social care research.}
\newblock {\em Public Health\/} .

\bibitem[\protect\citeauthoryear{Salganik and Heckathorn}{Salganik and
  Heckathorn}{2004}]{SalganikHeckathorn2004-RDS}
Salganik, Matthew~J. and Douglas~D. Heckathorn. 2004.
\newblock \enquote{Sampling and Estimation in Hidden Populations Using
  Respondent-Driven Sampling.}
\newblock {\em Sociological Methodology\/} 34:193--239.

\bibitem[\protect\citeauthoryear{Shelley, Bernard, Killworth, Johnsen, and
  McCarty}{Shelley et~al.}{1995}]{Shelley1995}
Shelley, Gene~A., H.~Russell Bernard, Peter Killworth, Eugene Johnsen, and
  Christopher McCarty. 1995.
\newblock \enquote{Who knows your {HIV} status? {W}hat {HIV}+ patients and
  their network members know about each other.}
\newblock {\em Social Networks\/} 17:189--217.

\bibitem[\protect\citeauthoryear{Traud, Kelsic, Mucha, and Porter}{Traud
  et~al.}{2011}]{facebook}
Traud, Amanda~L., Eric~D. Kelsic, Peter~J. Mucha, and Mason~A. Porter. 2011.
\newblock \enquote{Comparing Community Structure to Characteristics in Online
  Collegiate Social Networks.}
\newblock {\em SIAM Review\/} 53:526--543.

\bibitem[\protect\citeauthoryear{Traud, Mucha, and Porter}{Traud
  et~al.}{2012}]{facebooklong}
Traud, Amanda~L., Peter~J. Mucha, and Mason~A. Porter. 2012.
\newblock \enquote{Social structure of {Facebook} networks.}
\newblock {\em Physica A: Statistical Mechanics and its Applications\/}
  391:4165--4180.

\end{thebibliography}


\appendix
\appendixpage

Section \ref{appendix-derivations} details the derivations of the expectation and variance of the RR and RA estimators. Section \ref{aor-vs-roa} contains proofs that the one-step average-of-ratios degree estimators and prevalence estimators have greater variance than their ratio-of-averages counterparts under the binomial model. Section \ref{killworth-vs-ER} demonstrates that the binomial model approximates the Erd{\H o}s-R\'{e}nyi model; this is not a new result but is included here for completeness and convenience. Section \ref{dependence-on-other-params} examines the combinations of parameter values in which the RMSE is lower for the RA estimator than the RR estimator. Section \ref{zero-est-degrees} presents further details on the comparison of different methods for handling zero-valued degree estimates in the Facebook 100 simulations. Section \ref{additional-fb100-results} presents and discusses additional results from the Facebook 100 simulations.

\section{Derivations for RR and RA expectation and variance}
\label{appendix-derivations}

\[E(Y_{iH}) = \begin{cases}N_H p_{HH} & i\in H \\ N_H p_{HL} & i\in L \end{cases} \qquad E(Y_{iK}) = \begin{cases}N_K p_{HL} & i\in H \\ N_K p_{LL} & i\in L \end{cases}\]

\[\text{Var}(Y_{iH}) = \begin{cases}N_H p_{HH} (1-p_{HH}) & i\in H \\ N_H p_{HL}(1-p_{HL}) & i\in L \end{cases} \qquad \text{Var}(Y_{iK}) = \begin{cases}N_K p_{HL}(1-p_{HL}) & i\in H \\ N_K p_{LL}(1-p_{LL}) & i\in L \end{cases}\]

\[\sum_{i=1}^n E(Y_{iH}) = N_H \left(n_H p_{HH} + n_L p_{HL} \right) \qquad \sum_{i=1}^n E(Y_{iK}) = N_K \left(n_H p_{HL} + n_L p_{LL} \right)\]
\[\sum_{i=1}^n \text{Var}(Y_{iH}) = N_H \big[ n_H p_{HH} (1-p_{HH}) + n_L p_{HL}(1-p_{HL})\big]\]
\[\sum_{i=1}^n \text{Var}(Y_{iK}) = N_K \big[ n_H p_{HL} (1-p_{HL}) + n_L p_{LL}(1-p_{LL})\big]\]

First-order Taylor approximations for the mean and variance of a ratio of random variables are given by $E(A/B) \approx E(A)/E(B)$ and, if $A$ and $B$ are independent, $\text{Var}(A/B) \approx [E(B)^2 \text{Var}(A) + E(A)^2 \text{Var}(B)]/[E(B)^4].$ For handling more than one probe group and allowing probe groups and the hidden group to overlap with each other, one can either assume approximate independence or include the covariance term in the variance approximation; for now we consider the case of a single probe group $K$ that is disjoint from $H$.

Taking the expectation over the SBM superpopulation for a given sample,

\vspace{-1em}
\begin{align*}
    E\left(\hat{r}_\text{RR} \right) &= E\left(\frac{N_K}{N} \frac{\sum_{i=1}^n Y_{iH}}{\sum_{i=1}^n Y_{iK}} \right) \\
    &\approx \frac{N_K}{N} \frac{E\left(\sum_{i=1}^n Y_{iH}\right)}{E\left(\sum_{i=1}^n Y_{iK}\right)} && \text{1st order Taylor approximation} \\
    &= \frac{N_K}{N} \frac{\sum_{i=1}^n E\left( Y_{iH}\right)}{\sum_{i=1}^n E\left( Y_{iK}\right)} \\
    &= \frac{N_K}{N} \frac{n_H N_H p_{HH} + n_L N_H p_{HL}}{n_H N_K p_{HL} + n_L N_K p_{LL}} \\
    &= \frac{N_H}{N} \frac{n_H p_{HH} + n_L p_{HL}}{n_H p_{HL} + n_L p_{LL}} \\
    &= r \frac{n_H p_{HH} + n_L p_{HL}}{n_H p_{HL} + n_L p_{LL}}.
\end{align*}

Similarly, for the RA estimator,

\vspace{-1em}
\begin{align*}
    E\left(\hat{r}_\text{RA} \right) &= E\left(\frac{N_K}{N} \frac1n \sum_{i=1}^n \frac{Y_{iH}}{Y_{iK}} \right) \\
    &= \frac{N_K}{N} \frac1n \sum_{i=1}^n E\left(\frac{Y_{iH}}{Y_{iK}} \right) \\
    &\approx \frac{N_K}{N} \frac1n \sum_{i=1}^n \frac{E\left(Y_{iH} \right)}{E\left(Y_{iK} \right)} \\
    &= \frac{N_K}{N} \left[ \frac{n_H}n \frac{N_H p_{HH}}{N_K p_{HL}} + \frac{n_L}n \frac{N_H p_{HL}}{N_K p_{LL}} \right] \\
    &= r \left[ \frac{n_H}n \frac{p_{HH}}{p_{HL}} + \frac{n_L}n \frac{p_{HL}}{p_{LL}} \right].
\end{align*}

If $n_H/n \to r$, as in simple random sampling without replacement, then

\vspace{-1em}
\begin{align*}
    E\left(\hat{r}_\text{RR} \right) &\to r \frac{ r p_{HH} + (1-r) p_{HL}}{r p_{HL} + (1-r) p_{LL}}, \\
    E\left(\hat{r}_\text{RA} \right) &\to r \left[ r \frac{p_{HH}}{p_{HL}} + (1-r) \frac{p_{HL}}{p_{LL}} \right].
\end{align*}

\vspace{-1em}
\begin{align}
    \text{Var}\left(\hat{r}_\text{RR} \right) &= \text{Var}\left(\frac{N_K}{N} \frac{\sum_{i=1}^n Y_{iH}}{\sum_{i=1}^n Y_{iK}} \right) \nonumber \\
    &= \frac{N_K^2}{N^2} \text{Var}\left( \frac{\sum_{i=1}^n Y_{iH}}{\sum_{i=1}^n Y_{iK}} \right) \nonumber \\
    &\approx \frac{N_K^2}{N^2} \cdot \frac{E(\sum_{i=1}^n Y_{iK})^2 \text{Var}(\sum_{i=1}^n Y_{iH}) + E(\sum_{i=1}^n Y_{iH})^2 \text{Var}(\sum_{i=1}^n Y_{iK})}{E(\sum_{i=1}^n Y_{iK})^4} && \label{taylor} \\
    &= \frac{N_K^2}{N^2} \cdot \frac{\left(\sum_{i=1}^n E[Y_{iK}]\right)^2 \sum_{i=1}^n \text{Var}(Y_{iH}) + \left(\sum_{i=1}^n E[Y_{iH}]\right)^2 \sum_{i=1}^n \text{Var}(Y_{iK})}{\left(\sum_{i=1}^n E[Y_{iK}]\right)^4} && \label{indep} \\
    &= \frac{N_K^2}{N^2} \bigg( \frac{N_K^2 \left(n_H p_{HL} + n_L p_{LL} \right)^2 N_H \left[ n_H p_{HH} (1-p_{HH}) + n_L p_{HL}(1-p_{HL})\right]}{N_K^4 \left(n_H p_{HL} + n_L p_{LL} \right)^4} \:+\: \nonumber \\
    &\qquad\quad \frac{N_H^2 \left(n_H p_{HH} + n_L p_{HL} \right)^2 N_K \left[ n_H p_{HL} (1-p_{HL}) + n_L p_{LL}(1-p_{LL})\right]}{N_K^4 \left(n_H p_{HL} + n_L p_{LL} \right)^4} \bigg) \nonumber \\
    &= r \bigg( \frac1N \frac{\left(n_H p_{HL} + n_L p_{LL} \right)^2 \left[ n_H p_{HH} (1-p_{HH}) + n_L p_{HL}(1-p_{HL})\right]}{\left(n_H p_{HL} + n_L p_{LL} \right)^4} \:+\: \nonumber \\
    &\qquad\quad \frac{r}{N_K}\frac{ \left(n_H p_{HH} + n_L p_{HL} \right)^2 \left[ n_H p_{HL} (1-p_{HL}) + n_L p_{LL}(1-p_{LL})\right]}{\left(n_H p_{HL} + n_L p_{LL} \right)^4} \bigg). \nonumber 
\end{align}

Step (\ref{taylor}) above uses the Taylor approximation and Step (\ref{indep}) holds if $Y_{iG}, Y_{jG}$ are independent for $i\ne j$, $G = H,K$.  A similar derivation for the RA estimator yields

\vspace{-1em}
\begin{align*}
    \text{Var}\left(\hat{r}_\text{RA} \right) &\approx \frac{r}{n^2 N p_{HL}^2} \big[ n_H p_{HH}(1-p_{HH}) + n_L p_{LL}(1-p_{HL}) \big] \:+ \\
    &\qquad \frac{r^2}{n^2 N_K p_{HL}^3} \big[ n_H p_{HH}^2(1-p_{HL}) + n_L p_{HL}^2(1-p_{LL}) \big].
\end{align*}
\vspace{-1em}

If $n_H/n \to r$, then

\vspace{-1em}
\begin{align*}
    \text{Var}\left(\hat{r}_\text{RR} \right) &\to \frac{r}{nN} \frac{\left(r p_{HL} + (1-r) p_{LL} \right)^2 \left[ r p_{HH} (1-p_{HH}) + (1-r) p_{HL}(1-p_{HL})\right]}{\left(r p_{HL} + (1-r) p_{LL} \right)^4} \:+\: \\
    &\qquad\quad \frac{r^2}{nN_K}\frac{ \left(r p_{HH} + (1-r) p_{HL} \right)^2 \left[ r p_{HL} (1-p_{HL}) + (1-r) p_{LL}(1-p_{LL})\right]}{\left(r p_{HL} + (1-r) p_{LL} \right)^4} \\
    \text{Var}\left(\hat{r}_\text{RA} \right) &\to \frac{r}{n N p_{HL}^2} \big[ r p_{HH}(1-p_{HH}) + (1-r) p_{LL}(1-p_{HL}) \big] \:+ \\
    &\qquad \frac{r^2}{n N_K p_{HL}^3} \big[ r p_{HH}^2(1-p_{HL}) + (1-r) p_{HL}^2(1-p_{LL}) \big].
\end{align*}

Note that whether the link probabilities are constant as a function of $N$ (as under the constant density assumption, so that the average degree grows linearly with total network size $N$) or scaled by $1/N$ (so that average degree is constant with respect to $N$), the expressions for expectation and variance derived in this section will not change because the additional factors of $N$ cancel.

\section{Comparing the variances of A and R estimators}
\label{aor-vs-roa}

The two one-step degree estimators under consideration are as follows: \[\hat{D}_{i,\text{R}} = N\cdot\frac{\sum_j Y_{ij}}{\sum_j N_j}, \quad \hat{D}_{i,\text{A}} = N\cdot\frac1k \sum_{j=1}^k \frac{Y_{ij}}{N_j}.\] In the unlikely case that the probe groups are all the same size, then the two estimators are identical and therefore also have the same variance. We explore how the variances compare outside of this special case. Under the binomial model, $Y_{ij} \sim \text{Binom}(D_i,\: N_j/N)$. Therefore, \[\text{Var}\left(\hat{D}_{i,\text{R}}\right) = D_i \left[ N\cdot \frac1k \frac1{\sum_j N_j / k} - \frac{\sum_j N_j^2}{\left(\sum_j N_j\right)^2} \right].\] The second term within the brackets is less than $1/k$ because $N_j > 1$ for each probe group $j$. The first term equals $1/k$ if the average probe group size is equal to $N$. Therefore, the second term is negligible as long as the average probe group size is small compared to $N$: \[\text{Var}\left(\hat{D}_{i,\text{R}}\right) \approx D_i \left[ N\cdot \frac1k \frac1{\sum_j N_j / k} \right] = \frac{D_i N}k \frac1{\text{mean}(N_j)}.\] Following similar reasoning, \[\text{Var}\left(\hat{D}_{i,\text{A}}\right) = D_i \left[ \frac{N}k \frac1k\frac1{\sum_j N_j} - \frac1k \right] \approx \frac{D_i N}k \frac1k\frac1{\sum_j N_j} = \frac{D_i N}k \frac1{\text{hmean}(N_j)},\] where hmean$(N_j)$ denotes the harmonic mean of the probe group sizes. Since $N_j > 0$ for all $j$, the harmonic mean is strictly smaller than the arithmetic mean, and therefore $\text{Var}\left(\hat{D}_{i,\text{R}}\right) < \text{Var}\left(\hat{D}_{i,\text{A}}\right)$ if the average probe group size is sufficiently smaller than $N$.

A similar but simpler argument demonstrates the same result for the two one-step prevalence estimators (with fixed or known degrees) by comparing the arithmetic and harmonic means of the degrees: \[\hat{r}_{\text{R}} = \frac{\sum_i Y_{iH}}{\sum_i D_i},\quad \hat{r}_{\text{A}} = \frac1n \sum_i \frac{Y_{iH}}{D_i}.\] Again, we ignore the trivial and impractical case in which all degrees are identical. Under the binomial model, $Y_{iH} \widesim[2]{\text{indep}} \text{Binom}(D_i, \:r)$. Therefore, \[\text{Var}\left(\hat{r}_{\text{R}}\right) = \frac{r(1-r)}{n} \cdot \frac1{\sum_i D_i/n} = \frac{r(1-r)}{n} \cdot \frac1{\text{mean}(D_i)},\] \[\text{Var}\left(\hat{r}_{\text{A}}\right) = \frac{r(1-r)}{n} \cdot \frac1n \sum_i \frac1{D_i} = \frac{r(1-r)}{n} \cdot \frac1{\text{hmean}(D_i)}.\] For $D_i > 0$, outside of the case that all degrees are equal, the harmonic mean is strictly smaller than the arithmetic mean of the degrees. Therefore, $\text{Var}\left(\hat{r}_{\text{R}}\right) < \text{Var}\left(\hat{r}_{\text{A}}\right)$.

\section{Relating the binomial and Erd{\H o}s-R\'{e}nyi models}
\label{killworth-vs-ER}

Given a set of nodes, a graph can be simulated from an Erd{\H o}s-R\'{e}nyi model by conducting an iid Bernoulli trial $\ell_{ij}$ for each pair of nodes to determine whether there is a link between them: \[\ell_{ij} \stackrel{\text{iid}}{\sim} \text{Bernoulli}(p)\] Here we will suppose that the nodes have group memberships (either hidden or not hidden), but that these memberships do not impact link formation.

We can then derive random variables for the number of people each person $i$ knows who are in the hidden group: \[y_i = \sum_{j\in H, j\ne i} \ell_{ij} \sim \begin{cases}\text{Binom}\left(N_H, \:\:p\right) & i\notin H, \\ \text{Binom}\left(N_H-1, \:\:p\right) & i\in H, \end{cases}\]

and the number of people each person $i$ knows who are not in the hidden group: \[z_i = \sum_{j\notin H, j\ne i} \ell_{ij} \sim \begin{cases}\text{Binom}\left(N-N_H-1, \:\:p\right) & i\notin H, \\ \text{Binom}\left(N-N_H, \:\:p\right) & i\in H. \end{cases}\]

Let $N_H^* = N_H$ if $i \notin H$ and $N_H - 1$ if $i \in H$. Let $N_L^* = N-N_H-1$ if $i \notin H$ and $N-N_H$ if $i \in H$. Note that $N_H^* + N_L^* = N-1$ in either case, so $N_L^* = N- N_H^*-1$. The distributions also simplify notationally to \[y_i \sim \text{Binom}\left(N_H^*, \:\:p\right), \quad z_i \sim \text{Binom}\left(N_L^*, \:\:p\right).\]

These two variables are independent of one another for a given person, and their sum is that person's degree: \[d_i = y_i + z_i.\]

Consider two people $i$ and $j$. Their degrees are not completely independent because there is one potential link between them, so their degrees (sums of potential links) each include $\ell_{ij}$. Additionally, their responses are not independent if they are both in the hidden group, because in that case $y_i$ and $y_j$ both correspond to sums that include $\ell_{ij}$. Therefore, their conditional responses $y_i | d_i$ are not strictly independent of one another. However, for sufficiently large $N$ and $N_H$ this departure from independence is negligible.

The following derivation demonstrates that $y_i | d_i$ follows a hypergeometric distribution:

\vspace{-1.5em}
\begin{align*}
    p(y_i = y \:|\: d_i = d) &= \frac{p(y_i = y, d_i = d)}{p(d_i = d)} \\
    &= \frac{p(y_i = y, z_i = d-y)}{p(d_i = d)} \\
    &= \frac{p(y_i = y) p(z_i = d-y)}{p(d_i = d)} && y_i, z_i \text{ indep} \\
    &= \frac{p(y_i = y) p(z_i = d-y)}{\sum_{k=0}^{d_i} p(d_i = d \:|\: y_i = k) p(y_i = k)} \\
    &= \frac{p(y_i = y) p(z_i = d-y)}{\sum_{k=0}^d p(y_i = k) p(z_i = d-k)}
\end{align*}

\vspace{-1.5em}
\begin{align*}
    p(y_i = y) p(z_i = d-y) &= \binom{N_H^*}{y} \binom{N_L^*}{d-y} p^y (1-p)^{N_H^* - y} p^{d-y} (1-p)^{N_L^* - d+y} \\
    &= \binom{N_H^*}{y} \binom{N_L^*}{d-y} p^d (1-p)^{N-1-d}
\end{align*}

The only dependence on $y$ appears in the binomial coefficients: \[p(y_i = y \:|\: d_i = d) \propto \binom{N_H^*}{y} \binom{N_L^*}{d-y}.\] Therefore, $y_i|d_i$ follows a hypergeometric distribution for the number $y$ of successes out of $d$ draws without replacement from a population of size $N-1$ containing $N_H^*$ many successes.

For sufficiently large $N$ and $N_H$, the hypergeometric distribution of $y_i|d_i$ under the Erd{\H o}s-R\'{e}nyi model converges to the binomial distribution of $y_i|d_i$ under the binomial model, and the approximate independence of $y_i|d_i$ and $y_j|d_j$ for different people $i$ and $j$ converges to independence. In both models, the response $y_i$ can be interpreted as building person $i$'s personal network by drawing one person simply at random from the population and counting how many people were drawn that were in the hidden population; however, these draws are modeled without replacement under the Erd{\H o}s-R\'{e}nyi model and with replacement under the binomial model. Modeling without replacement seems more natural since people should not be double-counted in a given person's personal network.

The binomial model is an approximation for the conditional distribution of ARD responses in networks generated from an Erd{\H o}s-R\'{e}nyi model, and the approximation improves as $N$ and $N_H$ increase.

\section{Examining the dependence of comparative estimator performance on other parameters}
\label{dependence-on-other-params}

\begin{figure}
    \centering
    \includegraphics[width=0.9\textwidth]{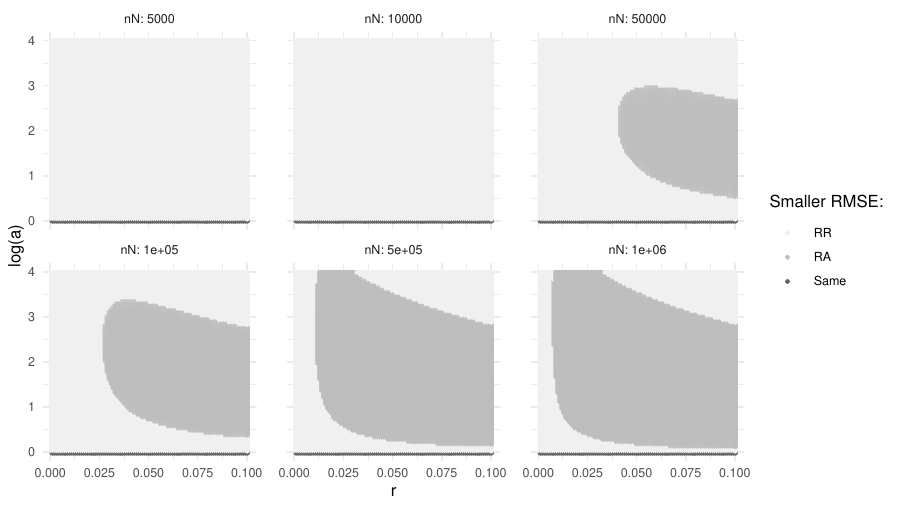}
    \caption{The six subpanels here correspond to six different values of $nN$, the product of the sample size and the population size. The size of the darkest region, the region in which the RA has smaller RMSE than the RR, grows with $nN$ (larger sample sizes and larger populations). All assortative and Erd{\H o}s-R\'{e}nyi simulations for each value of $nN$ are shown; $R$ ranges from 0.01 to 0.99, $\log(a)$ ranges from 0 to 4, $p_{HL}=0.01$, and $r_K$ ranges from 0.01 to 0.8.}
    \label{dependence-on-nN}
\end{figure}

\begin{figure}
    \centering
    \includegraphics[width=0.9\textwidth]{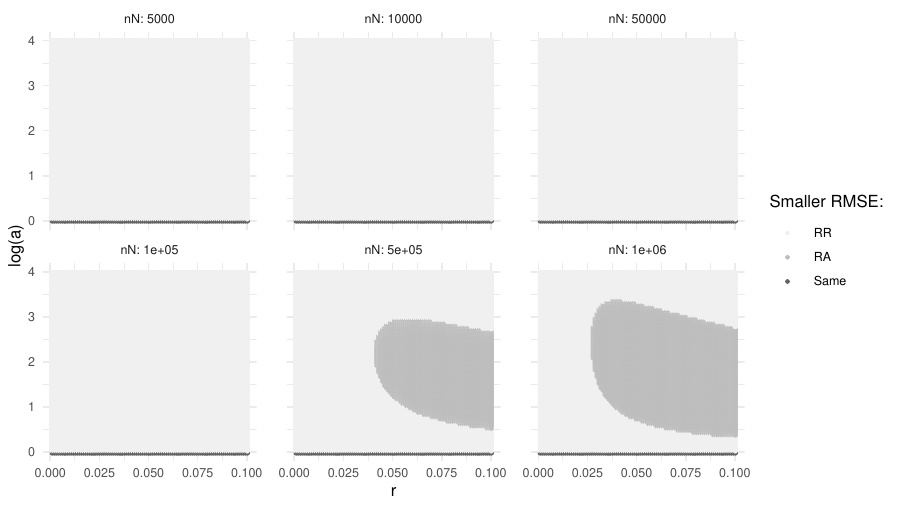}
    \caption{As in Figure \ref{dependence-on-nN} but with $p_{HL}=0.001$ instead of 0.01. The size of the region in which the RA has smaller RMSE than the RR is smaller for smaller $p$.}
    \label{dependence-on-p}
\end{figure}

\begin{figure}
    \centering
    \includegraphics[width=\textwidth]{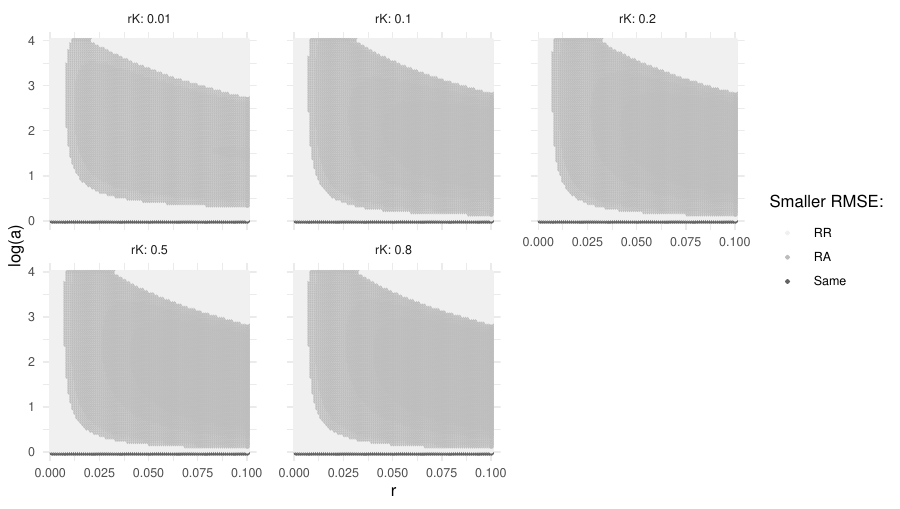}
    \caption{The size of the region in which the RA has smaller RMSE than the RR depends on $r_K$, the ratio of the probe group size to the population size, to a smaller extent than on $nN$. The five subpanels here correspond to five different values of $r_K$. All assortative and Erd{\H o}s-R\'{e}nyi simulations for each value of $nN$ are shown; $r$ ranges from 0.01 to 0.99, $\log(a)$ ranges from 0 to 4, $p_{HL}=0.01$, and $nN$ ranges from five thousand to one million.}
    \label{dependence-on-rK}
\end{figure}

Figure \ref{dependence-on-nN} examines the dependence of the region in which RA has lower RMSE on the value of $nN$, the product of the sample size and the population size, when $p=0.01$ and the other parameters are allowed to vary. For $nN$ as small as five or ten thousand, the RA no longer has lower RMSE. The size of the region increases with $nN$, expanding to encompass smaller values of $r$ and a wider range of $a$.

Figure \ref{dependence-on-p} shows the same results as Figure \ref{dependence-on-nN} except for a smaller value of $p$, 0.001 instead of 0.01. With smaller $p$, the region in which RA has smaller RMSE shrinks. Figure \ref{dependence-on-rK} shows that to a smaller extent, the size of this region also depends on the value of $r_K$, the ratio of probe group size to population size.

\section{Methods considered for handling zero-valued estimated degrees}
\label{zero-est-degrees}

We compute the RA and AA estimators four different ways to evaluate the impact of different methods for handling zero-valued degree estimates. Recall that the ratio $Y_{iH}/\hat{D}_i$ is NA when both the response $Y_{iH}$ and degree estimate $\hat{D}_i$ are 0, and infinite when only the degree estimate $\hat{D}_i$ is 0.

\begin{itemize}
    \item Approach 1: Set Infs to 1 and exclude NAs. This reduces the effective sample size by dropping people who have $Y_{iH}=\hat{D}_i = 0$.
    \item Approach 2: Set Infs to 1 and NAs to 0.
    \item Approach 3: Exclude Infs and NAs. This reduces the effective sample size by dropping all people with $\hat{D}_i = 0$.
    \item Approach 4: Exclude Infs and set NAs to 0. This reduces the effective sample size by dropping only people who have non-zero $Y_{iH}$ but $\hat{D}_i = 0$.
\end{itemize}

The reasoning behind setting NAs to 0 is that in the simulations, the numerator $Y_{iH}=0$ is known and without error, while the denominator $\hat{D}_i=0$ is estimated, and the true degree must be greater than 0. Therefore, even without knowing the true degree, the ratio is known to be 0. In real life, however, there would be error in both the response $Y_{iH}$ and the estimated degree $\hat{D}_i$. Therefore it seems Methods 1 and 3 might be the most reasonable to use in practice without further study.

Respondents with infinite-valued ratios have a non-zero response $Y_{iH}$ divided by a zero-valued degree estimate. In simulation world, the true denominator is known to be greater than zero but its exact value could make a big difference in the estimate, especially if $Y_{iH}$ is small. For example, if $Y_{iH} =1$, the ratio is 1/1 if the degree is 1 and 1/2 if the degree is 2. Should we actually set the Infs to NA and exclude them, computing the estimate with a reduced sample size? Degree estimates are more likely in the Facebook 100 data set to be zero when the true degree is small, and the true ratio tends to be close to 1 for these cases. These two properties could be likely in practice as well, but further research is needed to explore this.

The choice of method for handling zero-valued estimated degrees does not impact the overall findings of the paper; regardless of the method, the RA estimator has lower bias and RMSE than the other estimators in more cases than not, for instance. However, the choice of method sometimes affects which estimator has the lowest RMSE. For instance, under Method 1, the RA (AA, RR, AR) estimator has the lowest RMSE in 41\% (21\%, 20\%, 18\%) of the cases. From Method 1 to 4, the percentage increases for the RA and AA estimators and decreases for the RR and AR estimators, resulting in 43\% (24\%, 17\%, 16\%) for Method 4. This indicates that it changes the ranking of the estimators in some of the cases.

In this simulation study, then, we chose to show results using Method 1 since this seems like one of the more plausible methods in a real study.

\section{Additional results from the Facebook 100 simulations}
\label{additional-fb100-results}

\begin{figure}
    \centering
    \begin{subfigure}{0.45\textwidth}
        \includegraphics[width=\textwidth]{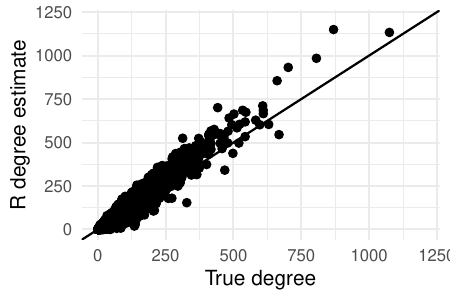}
        \caption{}
        \label{fig:first}
    \end{subfigure}
    \hfill
    \begin{subfigure}{0.45\textwidth}
        \includegraphics[width=\textwidth]{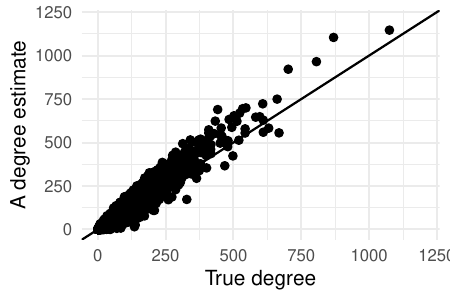}
        \caption{}
        \label{fig:second}
    \end{subfigure}
            
    \caption{Comparing the (a) R and (b) A degree estimates against the true degrees for a random sample of 10,000 people across the 100 school networks.}
    \label{fig:degree-estimates}
\end{figure}

\begin{figure}
    \centering
    \includegraphics[width=\textwidth]{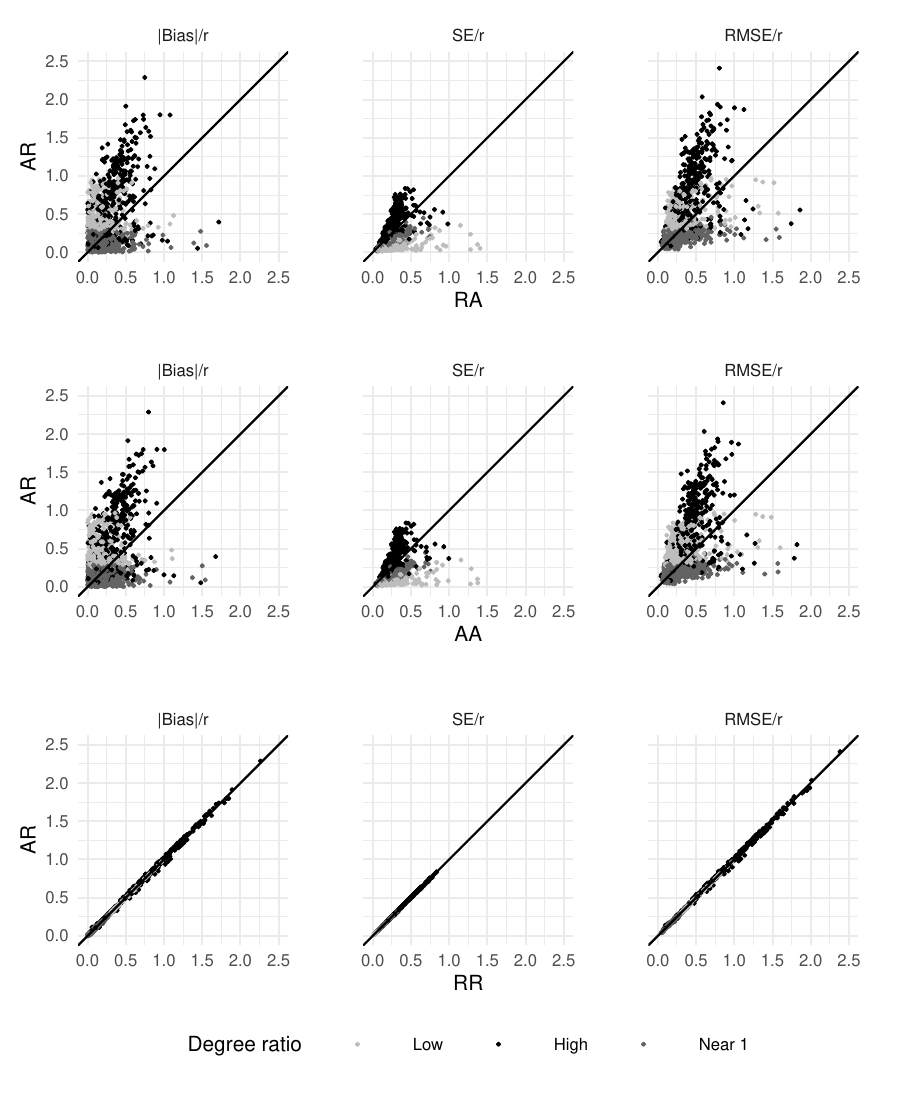}
    \caption{Comparing the absolute-value bias (left panel), standard error (center panel), and RMSE (right panel), all standardized by the true prevalence, of the AR estimator against that of the other three estimators in all 999 cases from the Facebook 100 simulations. The diagonal line is the one-to-one line; points above the line have \textbf{higher} values for the AR estimator than the other estimator. Each point represents the average across 500 surveys of size 500 for one combination of school network and hidden group.}
    \label{compare-with-AR}
\end{figure}

Figure \ref{fig:degree-estimates} presents scatterplots of the estimated degrees against the true degrees. The R and A estimates tend to be overestimates but are well correlated with the true degrees. Figure \ref{compare-with-AR} illustrates that the AR and RR estimators are comparable, and the RA and AA estimators tend to have lower bias and RMSE than the AR estimator for cases with low or high degree ratios.

\end{document}